\documentclass[final,5p,times,twocolumn]{elsarticle}

\usepackage{graphics}
\usepackage{amssymb}

\journal{Phys. Lett. A}

\newcommand{\be}{\begin{equation}}\newcommand{\ee}{\end{equation}}
\newcommand{\bea}{\begin{eqnarray}}\newcommand{\eea}{\end{eqnarray}}
\newcommand{\beaa}{\begin{eqnarray}}\newcommand{\eeaa}{\end{eqnarray}}
\newcommand{\ba}{\begin{array}}\newcommand{\ea}{\end{array}}
\newcommand{\bit}{\begin{itemize}}\newcommand{\eit}{\end{itemize}}
\newcommand{\ben}{\begin{enumerate}}\newcommand{\een}{\end{enumerate}}

\def\lf{\left}

\def\rar{\rightarrow}
\def\ri{\right}

\def\al{\alpha}
\def\Ga{\Gamma}\def\de{\delta}
\def\te{\theta}

\def\si{\sigma}
\def\om{\omega}

\def\1{{_{1}}}\def\2{{_{2}}}

\def\ZzZ{{\hbox{\tenrm Z\kern-.31em{Z}}}}
\def\CcC{{\hbox{\tenrm C\kern-.45em{\vrule height.67em width0.08em depth-
.04em \hskip.45em }}}}

\def\mapbelow#1{\smash{\mathop{\longrightarrow}\limits_{#1}}}

\newcommand{\lab}{\label}
\newcommand{\non}{\nonumber}
\newcommand{\bc}{\begin{center}}
\newcommand{\ec}{\end{center}}

\begin{document}

\begin{frontmatter}


\title{Rotating wave approximation and entropy}
\author[a1]{Andreas Kurcz}
\ead{andreas.kurcz@quantuminfo.org}
\author[a2]{Antonio Capolupo}
\ead{capolupo@sa.infn.it}
\author[a1]{Almut Beige}
\ead{a.beige@leeds.ac.uk}
\author[a3]{Emilio Del Giudice}
\ead{emilio.delgiudice@mi.infn.it}
\author[a2]{Giuseppe Vitiello}
\ead{vitiello@sa.infn.it}

\address[a1]{The School of Physics and Astronomy, University of Leeds, \\Leeds,
  LS2 9JT, United Kingdom}
\address[a2]{Dipartimento di Matematica e Informatica and \\I.N.F.N.,
  Universit\'a di Salerno, Fisciano (SA) - 84084, Italy}
\address[a3]{I.N.F.N. Sezione di Milano, Universit\`a di Milano, I-20133 Milano, Italy}

\begin{abstract}
This paper studies composite quantum systems, like atom-cavity systems and
coupled optical resonators, in the absence of external driving by resorting to
methods from quantum field theory. Going beyond the rotating wave
approximation, it is shown that the usually neglected counter-rotating part of
the Hamiltonian relates to the entropy operator and generates an irreversible
time evolution. The vacuum state of the system is shown to evolve into a
generalized coherent state exhibiting entanglement of the modes in which the
counter-rotating terms are expressed. Possible consequences at observational
level in quantum optics experiments are currently under study.
\end{abstract}

\begin{keyword}
quantized fields \sep quantum optics \sep rotating wave approximation \sep
entropy thermodynamics
\end{keyword}

\end{frontmatter}

\section{Introduction}

This article concerns one of the most useful approximations in quantum optics and atomic physics, namely the so-called rotating wave approximation (RWA) \cite{Gerry:2005a}. Up to now, there is in general very good agreement between experimental findings and theoretical predictions based on this approximation. Different reasons for the validity of the RWA are given in the literature. Most authors argue with time scale separation. Indeed, in the presence of sufficiently weak resonant interactions (like resonant laser driving of optical transitions) it is possible to move into an interaction picture, where the counter-rotating terms oscillate very rapidly. Their contribution to the time evolution of the system hence remains negligible when compared with the effect of the non-rotating terms \cite{Grynberg}. Other authors apply the RWA in order to preserve quantum numbers and energy \cite{Milonni,Schleich} and the validity of the so-called two-level approximation \cite{Meystre}.

However, recent trends in experimental quantum optics and atomic physics aim at the realisation of miniaturised devices \cite{Trupke,Reichel} with coupling constants which are many orders of magnitude larger than comparable, more classical designs \cite{Rempe}. As a result, the separation of the relevant time scales in a system might be reduced by many order of magnitude such that consequences of the normally neglected counter-rotating terms in the system Hamiltonian become observable. Systematic studies of the qualitatively different parameter regime, i.e.~beyond the rotating wave approximation, are currently becoming feasible in quantum optics experiments. An example, where the counter-rotating terms are already routinely taken into account, is the calculation of temperature limits in laser and in cavity cooling experiments \cite{review,NJPcooling}.

But there are other situations in which the RWA should not be applied. Also in certain situations far away from resonance, the counter-rotating terms should not be neglected \cite{Guo}. An example is discussed in Ref.~\cite{hegerfeldt2} by Hegerfeldt, who showed that the interaction between two atoms and the free radiation field, when treated exactly, can result in a small violation of Einstein's causality. Zheng {\em et al.}~\cite{Zubairy} also avoided the RWA and predicted corrections to the spontaneous decay rate of a single atom at very short times. Recently, Werlang {\em et al.}~\cite{Werlang} and the current authors \cite{concentration} pointed out that it might be possible to obtain photons by simply placing atoms inside an optical cavity. A similar energy concentrating effect might contribute significantly to the observed very high temperatures in sonoluminescence experiments \cite{sonolumi}. 

What Refs.~\cite{review,NJPcooling,hegerfeldt2,Zubairy,Werlang,concentration,sonolumi} have in common is that all of them consider {\em open} quantum systems, i.e. systems coupled with environment or with other systems (acting as environment). In such open systems energy is no longer necessarily a preserved quantity and time evolution is irreversible. For example, Refs.~\cite{Werlang,concentration} predict an energy concentrating mechanism in coupled atom-cavity systems, even in the absence of external driving.  This might seem unphysical but does not constitute a violation of the laws of thermodynamics, as long as the predicted changes in the free energy are accompanied by respective changes in entropy. 

\begin{figure}[t]
\resizebox{\columnwidth}{!}{\rotatebox{0}{\includegraphics{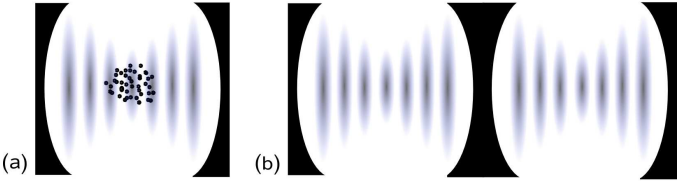}}}
\caption{Schematic views of two possible experimental realisations of two linearly-coupled bosonic reservoirs. (a) The experimental setup considered here could consist of a large number of strongly-confined two-level atoms inside the free radiation field or inside a relatively large optical resonator with a finite bandwidth. (b) The same Hamiltonian is obtained for two coupled optical resonators with overlapping field modes.} \label{fig1}
\end{figure}

In this paper, we consider a composite quantum system which is much simpler but nevertheless closely related to the one considered in Refs.~\cite{Werlang,concentration}. Resorting to quantum field theory (QFT) methods, we show that the two linearly-coupled bosonic reservoirs shown in Fig.~\ref{fig1} are characterised by a non-trivial entropy operator caused by the coupling of field modes and generating an irreversible time evolution. We show that the origin of the dissipative dynamics are indeed the usually neglected counter-rotating terms in the interaction between different components of the system.

In particular, we study the meaning and the effects of the RWA from the perspective of the space of the states of the system under study and show that when dealing with quantum fields, namely with infinitely many degrees of freedom, as it is necessary in the study of open systems, the counter-rotating terms cannot be neglected. This can be understood in general terms by observing that, as far as one limits himself to systems with finite numbers of degrees of freedom, the von Neumann theorem in
quantum mechanics (QM) guarantees that  the representations of the canonical
(anti-) commutation rules (CCR) (the Hilbert spaces of the states) are unitarily equivalent and therefore physically equivalent \cite{VonNeumann:1931a,umezawa,Bogoliubov}.  Thus, in QM there is no room for non-unitary transformations. One does not need to wonder about the  choice of the representation in which the system dynamics, i.e. the Lagrangian or the Hamiltonian operator, is realized since all of them have the same physical content; they are, indeed, equivalent up to a unitary transformation.

A different situation occurs when one deals with infinitely many degrees of freedom, as it happens when field operators are considered, i.e.~in QFT. In such a case, the von
Neumann theorem does not hold anymore and infinitely many unitarily
inequivalent representations of the CCR exist
\cite{VonNeumann:1931a,umezawa,Bogoliubov}. In this case, the choice which one
among these representations to adopt, i.e.~where to realize the system dynamics,
might be of crucial physical relevance. For example, one might realize the system
dynamics in a representation where the symmetry properties of the ground state
are the same as those of the field equations, or one might instead use a representation
where spontaneous symmetry breakdown occurs \cite{umezawa,Itzykson}. Also,   one might choose  a representation where the ground state is preserved under time evolution, or, instead, a representation  where it changes (as in unstable systems) under time translation transformations \cite{Celeghini:1992yv}; and so on.

The formal and physical significance of the unitarily inequivalence among
 representations is that the vacuum state in each of them cannot be expressed
 in terms of the vacua of other representations. Thus, for example, the
 vacuum of a metal in the superconductive phase cannot be expressed in terms
 of the vacuum of the (same) metal in the ``normal" phase. In
 phenomena related with unitarily inequivalence, a dominant role is typically played
 by ``weak" couplings  (e.g.~in the $\lambda \phi^4$ field theories, the order
 parameter, say $a$, which specifies the vacuum, is given by $a^2 \propto
 |m^2|/ \lambda$, where $\lambda$ is the coupling constant and $m^2$ is the
 (negative) squared mass). These weak coupling effects cannot be studied
 in a perturbative expansion around the vanishing value of the
 coupling constant (e.g.~in the $\lambda \phi^4$ model, the order parameter $a$
cannot be defined at $\lambda = 0$). As we will show, deciding whether or not to apply the RWA can constitute a similar delicate type of problem which, as said, cannot always be ignored when dealing with quantum fields.

Our calculations in the interaction picture reveal the underlying degrees of freedom which are involved in the generation of entropy and non-unitary time evolution in the setup shown in Fig.~\ref{fig1}. We show that the initial vacuum state of the system, i.e.~the state with no population in either mode, evolves in time into a generalized  SU(1,1) coherent state which is orthogonal to the initial state and exhibits entanglement of the modes in which the counter-rotating terms are expressed. The representation of the system then requires a different Hilbert space at any moment in time. This constant changing from one state space into another makes the dynamics of the system irreversible. When applying the RWA, the time evolution of the system remains unitary and the vacuum state of the system is always the same.

There are five sections in this paper. The theoretical model is introduced in Section \ref{Model}. In Section \ref{evolve} we show that the counter-rotating terms in the Hamiltonian of this system are related to the entropy operator and a detailed proof of irreversible time evolution is presented. In Section \ref{thermo} we study the free energy and the entanglement of the vacuum state. We show that the free energy is minimized at each time $t$ of the evolution of the vacuum state. Our discussion thus leads us to conclude that in conditions far off resonance the counter-rotating terms in the Hamiltonian are related to entropy, free energy and entanglement. These results are summarized in Section \ref{conc}. In the Appendix we report some mathematical formulas used in the text.

\section{Theoretical model and time-dependence of the vacuum state} \label{Model}

For the sake of simplicity and concreteness, it is convenient to refer to a specific model (which is widely used in quantum optics). Our discussion, however, can be extended to other models. We thus consider for example the familiar Hamiltonian $H$ for an ensemble of tightly-confined two-level systems (e.g.~atoms) coupled to a radiation field (cf.~Fig.~\ref{fig1}(a)). It can be written as \cite{Gerry:2005a}:
\be \lab{H}
 H = H_{ \rm 0} + H_{\rm int} ~ ,
\ee
where
\begin{eqnarray} \label{Hint}
H_0 = \sum _{{\bf k}} \hbar \omega _k ~ a _{\bf{k}} ^\dagger a
_{\bf k} + \sum _i {\frac{\hbar \omega _0}{2}} \sigma _{3i} ~ , \qquad&& \nonumber \\
H_{\rm int} =  i \hbar \sum _{\bf k} \sum_i g _{\bf k}  \left ( a _{\bf{k}} ^\dagger  - a _{\bf{k}} \right ) ( \sigma ^+ _i + \sigma ^- _i)~.&&
\end{eqnarray}
The $\sigma_i$'s are the two-level system SU(2) spin-like operators, $a_{\bf k}$ and $a_{\bf k} ^\dagger$  are the boson annihilation and creation operators of photon modes ${\bf k}$. The corresponding characteristic frequencies are $\omega_0$ and $\omega_k$. In the dipole approximation, the (real) atom-field coupling constants $g _{\bf k}$ in (\ref{Hint}) is given by  \cite{Gerry:2005a}
\begin{equation}
g _{\bf k} = - \left ( {\omega _k \over 2 \hbar \varepsilon _0 V} \right ) ^{1 / 2} ~ {\bf d}_{01} \cdot {\bf e} ~ ,
\end{equation}
where ${\bf d}_{01}$ is the dipole vector, ${\bf e}$ is the polarization vector, and $V$ is the volume. For simplicity we assume that the atoms are well localised within an optical domain and $e^{i {\bf k} {\bf r} _i} = 1$ for all wave vectors ${\bf k}$ and atomic positions ${\bf r}_i$. Moreover, all atoms have the same dipole moment ${\bf d}_{01}$. This is why we consider only one photon polarisation and why the $g_{\bf k}$ do not depend on $i$. Suppose $|0 \rangle_s$ denotes the vacuum for the two-level atoms, $|0 \rangle_r$ the vacuum state for the radiation field, and $|0 \rangle \equiv |0_s,0_r \rangle \equiv |0 \rangle_s \otimes |0 \rangle_r$. Then $\si^{-}|0 \rangle = 0 = a_{\bf k}|0 \rangle$.

Alternatively, one may consider a system of radiation field modes $a_{\bf k}$ and $a_{\bf k} ^\dagger$ interacting with another set of reservoir (or cavity) field modes with boson annihilation and creation operators $b_{\bf k}$ and $b_{\bf k}^\dagger$, as illustrated in Fig.~\ref{fig1}(b). The Hamiltonian $H$ can again be written as in (\ref{H}). Adopting a notation analogous to the one above, we now have $b _{\bf{k}}|0 \rangle = 0 = a_{\bf k}|0 \rangle$. The Hamiltonians $H_0$ and $H_{\rm int}$ are then given by
\begin{eqnarray} \lab{H0}
H_{ \rm 0} = \sum _{\bf k} \hbar \omega _k ~ a _{\bf k} ^\dagger a _{\bf k} + \sum
_{\bf k} \hbar \Omega _k ~ b _{\bf k} ^\dagger b _{\bf k} ~ , \hspace{0.28cm} &&\nonumber \\
H_{\rm int} = i \hbar \sum_{\bf k} g_{\bf k} ~ \lf(\, a^{\dag}_{\bf k}\, - \, a_{\bf k} \ri) \lf(b^{\dag}_{\bf k}\,+\,b _{\bf k} \ri) ~ ,&&
\end{eqnarray}
where the dipole approximation has been used again and where the $\Omega_k$ are the frequencies of the $b_{\bf k}$ modes.  Note that the (boson) operators $a_{\bf k}$ and $b_{\bf k}$ commute ($[a_{\bf k}, b_{\bf q}] = 0$, and all other commutators are zero). The commutation relations for them are
\bea \label{abcom}
[a_{\bf k},a_{\bf q}^\dag] = \de_{\bf k q} ~ ,    \qquad  [b_{\bf k},b_{\bf q}^\dag] = \de_{\bf k q} ~ .
\eea
All other commutators vanish.

It should be recalled here that the boson operators  $b_{\bf k}$ and $b_{\bf k}^\dagger$ can be related to the spin-like operators $\sigma^\pm$  by considering the ensemble of $N$ two-level systems (atoms) for large $N$. This is discussed in detail in Ref.~\cite{DeConcini:1976uk}. Since this derivation is outside the task of the present paper we omit to repeat it here again. We only recall that in the large $N$ limit ($N \rar \infty$) the Weyl-Heisenberg algebra is obtained as the contraction of the su(2) algebra for the $\si$'s \cite{DeConcini:1976uk}. As a result, in the large $N$-limit, both situations illustrated in Fig.~\ref{fig1} can be described by $H_{\rm 0}$ and $H_{\rm int}$ given in Eq.~(\ref{H0}). For definiteness, we consider these in the following.

Since we are here especially interested in the effect of the counter-rotating terms in the Hamiltonian $H_{\rm int}$, it is convenient to write
\begin{equation}
H_{\rm int} = H_{\rm JC} + H_{\rm CR}
\end{equation}
with
\begin{eqnarray}\lab{HI}
H_{\rm JC} = i \hbar \sum_{{\bf k}} g_{{\bf k} } ~ \lf(a^{\dag}_{{\bf k}}\,b_{\bf k}\,-\,a_{{\bf k}} \,b^{\dag}_{\bf k} \ri) ~ ,&& \nonumber \\
H_{\rm CR} = i \hbar \sum_{{\bf k}} g_{{\bf k}} ~\lf( a^{\dag}_{{\bf
    k}}\, b^{\dag}_{\bf k} \, -\, a_{{\bf k}} \, b_{\bf k}\ri) ~ .&&
\end{eqnarray}
The Hamiltonian $H_{\rm JC}$ is the usual Jaynes-Cummings Hamiltonian \cite{shore} and the part that usually survives in the RWA, while $H_{\rm CR}$ is the counter-rotating term part. There exists an interaction picture in which these counter-rotating terms are fast oscillating terms and they are ``therefore" neglected. The RWA consists in fact in neglecting those terms whose (antiresonant) frequencies ($\simeq \omega_{k,+} \equiv \omega_k + \Omega_k $) are far off from the resonance condition ($\omega_{k,-} \equiv \omega_k - \Omega_k \approx 0$).

From Eqs.~(\ref{H0}) and (\ref{HI}) we see that the vacuum $|0 \rangle$ is annihilated by $H_0$ and $H_{\rm JC}$ but not by $H_{\rm CR}$:
\be \lab{Hvac}
H_{\rm 0} \, |0 \rangle = 0, \qquad H_{\rm JC} \, |0 \rangle = 0, \qquad H_{\rm CR} \, |0 \rangle \neq 0 ~ .
\ee
This means that the vacuum of the theory is not invariant under time-translation unless the RWA is applied. This is an interesting feature from the mathematical point of view as well as from the physical point of view. It reminds us of the mechanism of spontaneous breakdown of symmetry. In the present case, the spontaneously broken symmetry is the one of time-translation: the dynamics is invariant under time-translation since obviously $[H,H] = 0$; however, the vacuum is not invariant since $e^{- i t H /\hbar} \, |0\rangle \neq |0\rangle$ due to Eq.~(\ref{Hvac}). So, let us consider this feature in more detail.

For notational simplicity, we focus our attention on only one of the ${\bf k}$ modes and omit the suffix ${\bf k}$ in the following. At the end we will recover them. In the Appendix we introduce the generators $J_i$ and $I_i$ $(i = 1,2,3)$ of the SU(1,1) and SU(2) group, respectively. In terms of these generators,  the Hamiltonian $H = H_0 + H_{\rm JC} + H_{\rm CR}$ with  $H_{0}$, $H_{\rm JC}$ and $H_{\rm CR}$ given by (\ref{H0}) and (\ref{HI}) can be written as
\bea \lab{HJI}
H =  \om_+ (J_3 - \frac{1}{2}) + \om_- I_3 - 2 g I_2 - 2g J_2
\eea
for each single $\bf k$ mode. For simplicity we have set $\hbar =1$. Notice that we have
\bea \lab{HRWJ}
H_{\rm CR} \propto    - 2 J_2 ~,
\eea
i.e. $H_{\rm CR}$  is, apart from the coupling factor, nothing but the $J_2$ generator of the SU(1,1) group (cf. Eq.~(\ref{J2})).

In order to study the effects of $H_{\rm CR}$ (cf. Eq.~(\ref{Hvac})) on the vacuum state, one could directly compute $e^{- i t H_{\rm CR}} \, |0\rangle$. However, it is instructive to see how the time evolution operator $e^{- i t H}$ itself, with $H$ given by Eq.~(\ref{HJI}), acts on the vacuum. To do that we rotate the state $e^{-itH} \, |0 \rangle$ into a frame which is more convenient for our study. This requires two successive rotations of $e^{-itH} \, |0 \rangle$ which are induced by the generators $I_1$ and $K_2$, respectively, with $K_2$ being the squeezing generator \cite{Yuen:1976vy} given in the Appendix (cf.~Eq.~(\ref{K2})). We thus consider the following rotations
\bea
\lab{IHtheta}
e^{-i t H} \, |0 \rangle &\longrightarrow & e^{i \theta I_{1}} e^{-i t H} \,
|0 \rangle = e^{-i t H(\theta)} \, |0 \rangle ~ , \nonumber \\
e^{-i t H(\theta)} \, |0 \rangle &\longrightarrow & e^{i \al K_{2}}e^{-i t
  H(\theta)} \, |0 \rangle = e^{-i t H (\theta,\al )} \, |0 (\al) \rangle ~.
\eea
The rotation angles $\theta$ and $\al$ are fixed by the theory parameters as
\be \lab{teta} \tan \theta = \frac{2g}{\om_-}~ , \qquad  \tanh \al = - \frac{4g^{2}}{\om_+ A}~ ,
\ee
respectively, with $A$ and $B$ given in the Appendix (cf.~Eq.~(\ref{Ad})). $H (\theta)$ and $H(\theta,\al)$ are given by $H (\theta) = e^{i \theta I_{1}} H~ e^{-i \theta I_{1}}$ and $H(\theta,\al) = e^{i \al K_{2}} H(\theta)~e^{-i \al K_{2}}$, respectively (cf. Eqs.~(\ref{IHteta}) and (\ref{KH})). We also have  $e^{i \theta I_{1}} \, |0 \rangle = |0 \rangle$. The state $|0 (\al) \rangle \equiv e^{i \al K_{2}} \, |0 \rangle$ is the squeezed vacuum \cite{Yuen:1976vy}.

In the corresponding interaction representation, the time evolution of the vacuum is solely
controlled by the interaction Hamiltonian in the interaction representation. There the $a$ and $b$ operators and their Hermitian conjugates carry their respective time dependence, $a(t)$ and $b(t)$ etc. Thus in the interaction representation with respect to the free Hamiltonian (cf. Eq.~(\ref{KH}))
\begin{equation}
\frac{B}{A}(J_3 - \frac{1}{2}) = {\cal E}(a^{\dag}a + b^{\dag}b)
\end{equation}
with ${\cal E} \equiv B/2A$, the interaction Hamiltonian $H^{\rm ip}_{\rm int}(\theta,\al)$ is (up to a constant term) given by
\bea \lab{KHip}
H^{\rm ip}_{\rm int}(\theta,\al) = \frac{\om_{+}A^{2}}{B} ~I_3 -  \frac{2g\om_{-}}{A} ~J_2   + 4g^2\frac{A}{B} ~K_1 ~ .
\eea
The operator $K_1$ is given in the Appendix (cf.~Eq.~(\ref{K21})). The operators $I_3$, $J_2$, and  $K_1$ have to be understood to be in the interaction representation (i.e.~they are made out of $a(t)$ and $b(t)$ etc.).

We now observe that
\bea
\lab{HipEv}
{}^{\rm ip}\langle 0(\al)|e^{-i t H^{\rm ip}_{\rm int} (\theta,\al )} |0 (\al)
\rangle^{\rm ip} \hspace{3.6cm} \\
\hspace{3.3cm} = \langle 0|  e^{-i \al K_{2}^{\rm ip}}e^{-i t H^{\rm ip}_{\rm int}
  (\theta,\al )}~ e^{i \al K_{2}^{\rm ip}}|0 \rangle \nonumber \\
\hspace{1cm}  = \langle 0|  e^{-i t {\tilde H}^{\rm ip}_{\rm int} (\theta,\al
  )}|0 \rangle~, \hspace{1.525cm}\nonumber
\eea
where
\be \lab{Hdiss1}
{\tilde H}^{\rm ip}_{\rm int} (\theta,\al ) = A~I_{3} (t) - 2 \Ga ~J_{2} (t)
\ee
with $\Ga \equiv g~\om_{-}/A$. Here we explicitly write $I_{3}(t)$ and $J_{2}(t)$ to remind the reader that we are using the interaction representation\footnote{It is interesting that the Hamiltonian of the
quantum damped oscillator has the same form as Eq.~(\ref{Hdiss1}). See Ref. \cite{Celeghini:1992yv}.}. Since $J_{2}(t)$ commutes with  $I_{3}(t)$  (cf. Eq.~(\ref{Cas})), and $I_{3}(t) \, |0  \rangle =0$, Eqs.~(\ref{HipEv}) and (\ref{Hdiss1}) yield
\bea
{}^{\rm ip}\langle 0(\al)|e^{-i t H^{\rm ip}_{\rm int} (\theta,\al )} |0 (\al) \rangle^{\rm ip} =
\langle 0|e^{i t 2 \Ga J_{2}(t) } |0 \rangle ~. \lab{Hvdiss}
\eea
By using
\be
e^{it{\cal E}(J_3 - \frac{1}{2})} e^{i t 2\Ga J_{2} (t)} e^{-it{\cal E}(J_3 - \frac{1}{2})} =
e^{i t 2\Ga J_{2}}
\ee
and
\be
e^{\pm i t{\cal E}(J_3 - \frac{1}{2})}|0\rangle = |0\rangle ~,
\ee
we finally obtain
\bea \lab{0t}
\langle 0|e^{i t 2\Ga J_{2}(t) } |0 \rangle = \langle 0|e^{i t 2\Ga J_{2} } |0 \rangle =\langle 0|0(t)\rangle ~.
\eea
Here the notation $|0(t)\rangle  \equiv e^{i t 2\Ga J_2 } |0 \rangle$ has been used. Note that $\Ga \rar 0$ in the resonant condition limit  $\om_{-} \rar  0$. From Eq.~(\ref{Hdiss1}) we see that the RWA is automatically implied in such a limit.  Eq.~(\ref{0t}) shows that the time evolution of the vacuum is in general non-trivial and indeed generated by an operator proportional to $J_2$.

In the next Section, we calculate an explicit expression for $\langle
0|0(t)\rangle$ by resorting to the well-known results of QFT
\cite{umezawa,Celeghini:1992yv,Perelomov:1986tf}. Our discussion will show
that the counter-rotating part of the system Hamiltonian, i.e.~$H_{\rm CR}$,
is related to irreversible time evolution, entropy, entanglement, and free
energy. To our knowledge, these properties of $H_{\rm CR}$ have so far been
ignored. In conditions far off the resonance or in the absence of external
driving they may produce observational consequences
\cite{concentration,sonolumi}.

\section{$H_{\rm CR}$, irreversible time evolution and entropy} \label{evolve}

In order to compute the quantity $\langle 0|0(t)\rangle$ explicitly, we start by denoting the set of simultaneous eigenvectors of $a^{\dag}a$ and $b^{\dag}b$ by $\{|n_{a},n_{b} \rangle \}$.  The corresponding eigenvalues $n_{a}$ and $n_{b}$ are non-negative integers \cite{Perelomov:1986tf}. Expressing
$I_{3}$ and $(J_{3} - \frac{1}{2})$, which in SU(1,1) form a complete set of commuting operators,  in this basis, we find
\bea
\lab{jm}
\hspace{1cm} I_{3} |j, m \rangle = j |j, m \rangle ~,~ \quad \hspace{0.2cm} j = \frac{1}{2}(n_{a} -
n_{b}) ~ ,&& \nonumber \\ \lab{jm2} \left(J_{3} - \frac{1}{2} \right) |j, m
\rangle = m |j, m \rangle~,~ \quad m= \frac{1}{2}(n_{a} + n_{b})
~. && 
\eea
Here $m \geq |j| \geq 0$ since $n_{a}$ and $n_{b}$ are non-negative \cite{Perelomov:1986tf}.  We remark that once  the eigenvalue of $I_{3}$ is set to be definite positive by boundary condition,  then it remains constant under the time evolution induced by $J_{2}$ (i.e. by $H_{\rm CR}$), as it must be  since $J_{2}$ and $I_{3}$ commute ($I_{3}$ is indeed the SU(1,1) Casimir operator). Also note that the original vacuum $|0 \rangle$ of the system is actually the eigenstate of $I_{3}$ and $(J_{3} - \frac{1}{2})$  associated with the zero eigenvalues $j=0$ and $m=0$, i.e. $|0 \rangle = | n_{a}=0,~ n_{b}= 0 \rangle$.

We are now ready to compute the explicit expression for $|0(t)\rangle$. By use of the ``normal form'' of the operator $e^{i t 2\Ga J_2 }$ (see e.g. chapter 4 of ref. \cite{Perelomov:1986tf} and refs. \cite{umezawa,Celeghini:1992yv}) we  obtain the  vacuum at time $t$
\be \lab{p79}
|0(t)\rangle \equiv e^{i t 2\Ga J_2 } |0 \rangle = {1\over{\cosh{(\Gamma  t)}}} \exp{ \left ( \tanh
{(\Gamma  t)} J_{+} \right )} |0 \rangle ~. 
\ee
At each time $t$, $|0(t)\rangle$ is a normalized state,
\bea \lab{norm}
 \langle 0(t)|0(t)\rangle  = 1~, \quad \forall t ~ .
\eea
In the limit $t \rar \infty$, the vacuum $|0(t)\rangle$ becomes orthogonal to the original vacuum $|0 \rangle = |0 (t = 0)\rangle $. Indeed we obtain:
\bea \lab{ort}
 \langle 0|0(t)\rangle  = \exp{(- \ln \cosh \Ga t)} ~~\mapbelow{t \rightarrow \infty} 0 ~.
\eea
The vacuum instability shown in Eq.~(\ref{ort}) has to be expected on the
basis of physical intuition since $H_{\rm CR}$, being related with fast
oscillating terms, introduces transient phenomena. These are of dissipative nature, their time evolution being controlled by $e^{-\Ga t}$ for large $t$,  as Eq.~(\ref{ort}) shows.

The time evolution induced by the counter-rotating terms in the Hamiltonian is thus shown to be only well defined for finite short time-intervals ($t < 1 / \Ga$). As $t \rar \infty$, the time evolution manifests itself as  a non-unitary transformation which  leads  out of the original Hilbert space whose vacuum is $|0 \rangle $. This is clearly a pathology, since due to the von Neumann theorem there is no room in QM for the non-unitary time evolution expressed by Eq.~(\ref{ort}). However, far off resonance (i.e.~for non-vanishing  $\om_{-}$), one cannot neglect the counter-rotating terms. In this case, it becomes unavoidable to study the system dynamics without performing the RWA.

We are thus led to explore the possibility to formulate our problem in the framework of QFT \cite{Celeghini:1992yv}, where infinitely many unitarily inequivalent representations exist. In order to do that, we first restore the suffices ${\bf k}$ and write $|0(t)\rangle$ as
\be |0(t)\rangle = \prod_{{\bf k}} {1\over{\cosh{(\Gamma_{k} t)}}}
\exp{ \left ( \tanh {(\Gamma_{k} t)} J_{+,{\bf k}} \right )} \, |0\rangle ~, \lab{pq79}\ee
where  $\Ga_{k} \equiv g_{\bf k} ~\om_{k, -} / A_{k}$. Eq.~(\ref{pq79})
is a formal relation holding for finite volume $V$.   As customary in QFT, one works at finite volume and the limit  $V \rightarrow \infty$ is taken only at the end of the computation. The state $|0(t)\rangle$ is again a normalized state, $\langle 0(t)|0(t)\rangle = 1$, at each time $t$. In fact, it is an SU(1,1) generalized coherent state \cite{Perelomov:1986tf}, i.e.~a two mode Glauber-type coherent state. Eq.~(\ref{ort}) is now replaced by
\be \langle 0 | 0(t)\rangle = \exp{\left ( - \sum_{{\bf k}} \ln \cosh
{(\Gamma_{ k} t)} \right )}  ~~\mapbelow{t \rightarrow \infty} 0 ~, \lab{pq711}\ee
which again exhibits non-unitary irreversible time evolution.

In QFT,  however, we have to consider infinitely many degrees of freedom. Thus, by using the continuous limit relation $ \sum_{{\bf k}} \mapsto
{V / (2 \pi)^{3}} \int \! d^{3}{{ k}}$, we obtain
\begin{eqnarray}
\langle 0 | 0(t)\rangle &\mapbelow{V \rightarrow \infty}& 0
\quad ~~ \forall \, t ~ , \lab{p713} \nonumber \\
\langle 0 (t') | 0(t)\rangle &\mapbelow{V \rightarrow \infty}& 0 \quad ~~\forall \, t\, , t' ~ {\rm with} ~ t
\neq t' ~, \lab{p714}
\end{eqnarray}
provided that $\int \! d^{3}{ k} \, \ln \cosh {(\Gamma_{ k} t)}$ is finite and positive.
The meaning of Eq.~(\ref{p713}) is that  the representation at a given time $t$ is unitarily inequivalent to the representation at any different time $t' \neq t$ in the infinite volume limit: The system spans a whole set of unitarily inequivalent representations as time evolves. Each of them is {\it labeled} by different values of $t$. The occurrence of such a phenomenon is possible in QFT where infinitely many unitarily inequivalent representations exist. Time evolution is thus described in terms of ``phase transitions'' among the representations, or ``trajectories'' in the  space of the representations.

The above calculations show that the generator of such a non-unitary
time evolution is the counter-rotating term proportional to $J_{2} \equiv \sum_{{\bf k}} J_{2,{\bf k}}$. We now show that $J_2$ is associated with the entropy operator.

The vacuum state $|0(t)\rangle$ given
by Eq. (\ref{pq79}) can also be written as \cite{umezawa,Celeghini:1992yv}
\be |0(t)\rangle \, = \exp{\left ( - {1\over{2}} {\cal
S}_{a} \right )} |\,{\cal I}\rangle \, = \exp{\left ( -
{1\over{2}} {\cal S}_{b} \right )} |\,{\cal I}\rangle
~ . \lab{p81}\ee
Here
\be |{\cal I}\rangle \equiv \exp {\left( \sum_{{\bf k}} a_{\bf k}^{\dagger} b_{\bf k}^{\dagger} \right)} |0\rangle  \lab{p821}\ee
is a not normalizable vector \cite{umezawa,Celeghini:1992yv} and
${\cal S}_{a}$ is given by
\be \lab{p831}
{\cal S}_{a}
\equiv - \sum_{{\bf k}} \Bigl \{ a_{{\bf k}}^{\dagger} a_{{\bf k}}
\ln \sinh^{2} \bigl ( \Gamma_{ k} t \bigr ) - {a_{{\bf k}}} a_{{\bf k}}^{\dagger} \ln \cosh^{2} \bigl ( \Gamma_{ k} t \bigr ) \Bigr
\} ~. 
\ee
${\cal S}_{b}$  has the same expression
with $b_{{\bf k}}$  and $b_{{\bf k}}^{\dagger}$
replacing $a_{{\bf k}}$  and $a_{{\bf k}}^{\dagger}$, respectively. In the following, we write ${\cal S}$  for either ${\cal S}_{a}$ or ${\cal S}_{b}$. It is not difficult to recognize that
${\cal S}$  is the entropy \cite{umezawa}.  Indeed, the state $|0(t)\rangle$ can be written in the form
\be \label{W}
|0(t)\rangle = \sum_{n=0}^{+\infty} \sqrt{W_n} |n,n \rangle
\ee
where $n$ denotes the set $\{n_k\}$ and
\be \label{wn}
  W_n (t) = \prod_k
  \frac{\sinh^{2n_k}( \Gamma_{ k} t \bigr )}{\cosh^{2(n_k +1)}( \Gamma_{ k} t \bigr )}
\ee
with
\be \label{wn1}
0< W_n <1 \quad {\rm and} \quad  \sum_{n=0}^{+\infty} W_n = 1 \,.
\ee
We have
\be \label{SW}
\langle 0 (t) |{\cal S}|0(t)\rangle = - \sum_{n=0}^{+\infty} W_n (t) \ln W_n (t)~.
\ee
Finally, for the time variation of $|0(t)\rangle$ at finite volume $V$, we
obtain
\be {{\partial}\over{\partial t}} |0(t)\rangle =  -
{1\over{2}} \left ( {{\partial {\cal S}}\over{\partial t}}
\right ) |0(t)\rangle \lab{p91}\ee
which shows that $ i (1 / {2}) \left ( {{\partial {\cal
S}}/ {\partial t}}  \right )$ acts as the generator of
time-translations. As observed elsewhere \cite{Celeghini:1992yv,DeFilippo:1977bk},
it is remarkable that the same dynamical variable ${\cal S}$ whose
expectation value is formally the entropy also controls time
evolution: A privileged direction in time evolution ({\it arrow
of time}) emerges which signals the breaking of time-reversal invariance.

\section{Free energy and entanglement} \label{thermo}

By acting with the operator $J_{2}$ on the operators $a_{{\bf k}}$ and $b_{{\bf k}}$ we obtain  the Bogoliubov transformations $a_{{\bf k}} \mapsto a_{{\bf k}}(\te(t))$, $b_{{\bf k}} \mapsto b_{{\bf k}}(\te(t))$:
\bea
a_{{\bf k}}(\te(t)) = {\it e}^{ i t 2 \sum_{{\bf q}} \Gamma_{ q} J_{2,{\bf q}}}  a_{{\bf k}}
{\it e}^{-i t 2 \sum_{{\bf q}} \Gamma_{ q} J_{2,{\bf q}}} \hspace{1cm}&& \\
= a_{{\bf k}} \cosh{\te_{ k}(t)} - b_{{\bf k}}^{\dagger} \sinh{\te_{ k}(t)}~, \hspace{0.5cm}
&& \non \\ \non \\
b_{{\bf k}}(\te(t)) = {\it e}^{ i t 2 \sum_{{\bf q}} \Gamma_{ q} J_{2,{\bf q}}}  b_{\bf k}
{\it e}^{-i t 2 \sum_{{\bf q}} \Gamma_{ q} J_{2,{\bf q}}} \hspace{1cm}&& \\
= - a_{{\bf k}}^{\dagger}
\sinh{\te_{ k}(t)} + b_{\bf k} \cosh{\te_{ k}(t)} ~, \hspace{0.225cm} \non
\eea
where $\te_{k}(t) \equiv \Gamma_{ k} t$. The $a_{{\bf k}}(\te(t))$ and $b_{{\bf k}}(\te(t))$ operators are the annihilation operators for the vacuum $|0(t)\rangle$ since
\be a_{{\bf k}}( \te(t))|0(t)\rangle  = 0 = b_{{\bf k}}( \te (t))|0(t)\rangle ~, \quad \forall \, t ~.
 \lab{p273}\ee
At each instant $t$ and for each ${\bf k}$, we have
\be {n}_{a_{\bf k}}(\te(t)) \equiv \langle 0(t) | a_{\bf k}^{\dagger}
a_{\bf k}| 0(t)\rangle  = \sinh^{2}\te_{ k}(t) ~ . \lab{p274}\ee
Similarly one obtains ${n}_{b_{\bf k}}(\te(t)) = \sinh^{2}\te_{ k}(t)$ for the $b_{{\bf k}}$ modes.

We have observed that the state $|0(t)\rangle$ is an SU(1,1) generalized coherent state. Eq.~(\ref{p274}) and the similar one for ${n}_{b_{\bf k}}( \te(t))$ show that it is a coherent condensate of equal number of $a_{\bf k}$ and $b_{\bf k}$ modes. One can show that the  creation of the $a_{\bf k}(\te(t))$ mode is equivalent to the  destruction of  the $b_{\bf k}(\te(t))$ mode and vice-versa \cite{umezawa,Celeghini:1992yv}. This means, the $b_{\bf k}(\te(t))$ modes can be interpreted as the {\it holes} for the $a_{\bf k}(\te(t))$ modes and vice-versa. In other words, the $b(\te(t))$-system can be considered as the sink where the energy dissipated by the $a(\te(t))$-system flows and vice-versa.

In this context, we also note that
\be [\, {\cal S}_{a} - {\cal S}_{b} ,  J_2 ] = 0~, \qquad [\, {\cal S}_{a} - {\cal S}_{b} ,  I_3 ] = 0
~. \lab{p88}\ee
Thus the difference ${\cal S}_{a} - {\cal S}_{b}$ is constant under the time evolution.
Since the $b(\te(t))$-modes are the holes for the $a(\te(t))$-modes, ${\cal S}_{a} - {\cal S}_{b}$ is in fact the entropy for the closed system.

By closely following  Ref. \cite{Celeghini:1992yv}
we introduce the free energy functional for the $a(\te(t))$-modes (we could do it as well for the $b(\te(t))$-modes)
\be F_{a} \equiv \langle 0(t)| \Bigl ( H_{0,a } -
 {1\over{\beta}} {\cal S}_{a} \Bigr ) |0(t)\rangle ~, \lab{p92}\ee
where $H_{0,a} \equiv \sum_{\bf k} E_{ k} a_{\bf
k}^{\dagger} a_{\bf k}$ with $E_{ k} \equiv  A_{ k}/2$. Assuming that $\beta =
{1 / {k_{B} T(t)}}$ ($k_{B}$ denotes the Boltzmann
constant) is a slowly varying function of the time $t$, the stability condition $\partial
F_{a} / \partial \theta_{ k} = 0$,  $\forall k,$  gives $\beta E_{ k} = - \ln \tanh^{2}  \theta_{ k} (t)$. Then
\be {n}_{a_{\bf k}}(\theta(t)) = \sinh^{2}\theta_{ k}(t) =
{1\over{e^{\beta (t) E_{ k}} - 1}} ~, \lab{p101} \ee
which is the Bose distribution for $a_{\bf k}$ at time $t$.
We thus recognize that  $\{ |0(t)\rangle \}$ is a representation
of the CCR at finite temperature \cite{umezawa}. We can show that
\be
d  F_{a} = d E_{k} -
{1\over{\beta}} d {\cal S}=0 ~.
\ee
Indeed we find
\be \lab{dq} d E_{k} = \sum_{\bf k}  \,E_{k} \,
\dot{n}_{a_{\bf k}}(\theta(t))d t = \frac{1}{\beta} d S = d Q ~.
\ee
Here  ${\dot{n}}_{a_{\bf k}}(\theta(t))$ denotes the time derivative of
${n}_{a_{\bf k}}(\theta(t))$, and, as usual, we define  heat as $dQ =
(1 / \beta) dS$.

We finally remark that the vacuum $|0 (t)\rangle$ can be written
in the following  form
\bea \label{ent}
|0(t) \rangle = \prod_k \frac{1}{\cosh \te_{k}(t)} [ |0 \rangle \otimes |0
\rangle \hspace{1.6cm}&& \\
+ \sum_{\bf k} \tanh \te_{k}(t) \left( | a_{\bf k} \rangle \otimes |b_{\bf k} \rangle
  \right)  + \dots ] \,. \non && 
\eea
This shows that it cannot be factorized into the product of two
single-mode states: it is an entangled
state for the  modes $a_{\bf k}$ and $b_{\bf k}$.
Eqs.~(\ref{W})-(\ref{SW}) then show that ${\cal S}$  provides a measure of
the degree of entanglement: the probability of having entanglement of the two sets of
$n$ $a$-modes  and $n$ $b$-modes  is $W_n$. Since
$W_n$ is a decreasing monotonic function of $n$, the
entanglement is suppressed for large $n$. It appears then, that
only a finite number of entangled terms in the expansion
(\ref{W}) is relevant. However, this is only true at
finite volume. The entanglement is truly realized in the infinite
volume limit, i.e. in QFT, where the summation in Eq.~(\ref{W}) extends to an infinite number of components and  Eq.~(\ref{p713}) holds \cite{Iorio}.

Notice that the robustness of the entanglement is
rooted in the fact that, once the infinite volume limit is reached,
there is no unitary generator able to disentangle the $a$ and $b$ modes.
Such a non-unitarity is only realized when all the terms in the
series (\ref{W}) are summed up, which indeed happens in
the $V\to \infty$ limit \cite{Celeghini:1992yv,Iorio}.

\section{Conclusions} \label{conc}

The discussion presented in this paper leads us to conclude that, provided the resonance condition $\om_- \approx 0$ does not apply,  the counter-rotating part $H_{\rm CR}$ of the Hamiltonian of two linearly coupled bosonic reservoirs shown in Fig.~\ref{fig1} may reveal very interesting dynamical features which are typical of non-perturbative quantum field theory. By resorting to known results \cite{umezawa,Celeghini:1992yv}, the discussion of the explicit expression  of $\langle 0|0(t)\rangle$ in such a frame for a concrete quantum optical system has shown that the RWA cannot be applied in conditions far off resonance. There are absolutely non-trivial features in the physical behavior of this system which have been overlooked so far to our knowledge.

In particular, for non-vanishing $\om_-$, $H_{\rm CR}$ turns out to be related to the entropy operator. This signals irreversible (non-unitary) time evolution, which is a manifestation of the breakdown of time-reversal invariance (the arrow of time). Time evolution of the system shown in Fig.~\ref{fig1} hence needs to be described in terms of ``trajectories'' in the  space of the unitarily inequivalent representations $\{|0(t) \rangle \}$. The free energy functional is minimized on such trajectories (at each time $t$ we have $dF = 0$).  In each representation, the system
ground state $|0(t) \rangle$ turns out to be a generalized coherent state which is an entangled state of the modes in terms of which $H_{\rm CR}$ is expressed.

Beyond their theoretical interest, our results may have some relevance also from the experimental standpoint \cite{concentration,sonolumi}. For example, a study of a possible energy concentrating mechanism in atom-cavity system, which might become the object of an actual experimental observation, is currently in progress with preliminary results reported in Ref.~\cite{concentration}.

\section{Acknowledgements}
A. B. acknowledges a James Ellis University Research Fellowship from the Royal Society and the GCHQ. This work was supported by the UK Research Council EPSRC, the University of Salerno, and INFN.

\appendix
\setcounter{section}{1}
\section*{Appendix}

We present below some formulas used in the derivations in the text. For simplicity
we omit the momentum suffices ${\bf k}$. Let us start by presenting the generators of the SU(1,1) group:
\bea \lab{J}
J_1 = \frac{1}{2}(J_+ + J_-) = \frac{1}{2}(a^\dag b^\dag + ab) ~ , \hspace{0.4cm}&&\\ \lab{J2}
J_2 = -\frac{i}{2}(J_+ - J_-) = -\frac{i}{2}(a^\dag b^\dag - ab) ~ ,&&\\ \lab{J3}
J_3 = \frac{1}{2}(a^\dag a + b^\dag b + 1) ~ ,\hspace{1.9cm}&&
\eea
or else
\bea \lab{Jpm}
J_+ = a^\dag b^\dag ~ , \qquad J_- =  ab ~ ,
\eea
with commutators
\bea \lab{Jcom}
[J_+,J_- ] = -2 J_3 ~ , \qquad  [J_3,J_\pm ] = \pm J_\pm ~ .
\eea
Notice that when the suffix ${\bf k}$ is introduced, we may define
$J_i = \sum_{\bf k} J_{\bf k i}~$, $i=1,2,3$, and then the group structure is
the one of $SU(1,1) = \bigotimes_{\bf k} SU_{\bf k}(1,1)$. We also introduce the generators of the SU(2) group:
\bea \lab{JA}
I_1 = \frac{1}{2}(I_+ + I_-) = \frac{1}{2}(a^\dag b + b^\dag a) ~ ,
\hspace{0.45cm} &&\\
I_2 = -\frac{i}{2}(I_+ - I_-) = -\frac{i}{2}(a^\dag b - b^\dag a) ~ , &&\\
I_3 = \frac{1}{2}(a^\dag a - b^\dag b) ~ , \hspace{2.35cm} &&
\eea
and
\be \lab{JpmA}
I_+ = a^\dag b , \qquad I_- =  b^\dag a ~ ,
\ee
with commutators
\bea \lab{JcomA}
[I_+,I_- ] = 2 I_3 ~ , \qquad  [I_3,I_\pm ] = \pm I_\pm ~ .
\eea
Notice that the Casimir operator for the $J$'s su(1,1) algebra  is $I_3$, while the Casimir operator for the $I$'s su(2) algebra  is $J_3 - \frac{1}{2} = \frac{1}{2}(a^\dag a + b^\dag b)$, i.e.
\bea \lab{Cas}
[J_3 - \frac{1}{2},I_i ] = 0 ~ , \qquad  [I_3,J_i ] = 0 ~ , \quad i = \pm 3 ~ .
\eea
The ${\tilde K}$'s and $K$'s generators given below also close the su(1,1) algebra:
\bea \lab{K1til}
{\tilde K}_1 = \frac{1}{4}[(a^2 + {a^\dag}^2) + (b^2 + {b^\dag}^2) ] ~ ,&&\\
{\tilde K}_2 = \frac{i}{4}[(a^2 - {a^\dag}^2) - (b^2 - {b^\dag}^2) ] ~ ,&&\\
{\tilde K}_3 = \frac{1}{2}[{a^\dag} a - {b^\dag} b ] = I_{3} ~ . \hspace{1.1cm}&&
\eea
with commutators
\bea \lab{Ktilcom}
[{\tilde K}_1 , {\tilde K}_2 ] = -i {\tilde K}_3 ~ , \qquad  [{\tilde K}_3 ,{\tilde K}_{1,2} ] = \pm i {\tilde K}_{2,1} ~ .
\eea
and
%
\bea \lab{K21}
K_1 = \frac{1}{4}[(a^2 + {a^\dag}^2) - (b^2 + {b^\dag}^2) ] ~ ,&&\\
\lab{K2}
K_2 = \frac{i}{4}[(a^2 - {a^\dag}^2) + (b^2 - {b^\dag}^2) ] ~ ,&&\\
K_3 = \frac{1}{2}[{a^\dag} a - {b^\dag} b ] = {\tilde K}_3 = I_{3} ~ . \hspace{0.31cm}&&
\eea
with commutators
%
\bea \lab{Kcom}
[K_1 , K_2 ] = -i K_3 ~ , \qquad  [K_3 ,K_{1,2} ] = \pm i K_{2,1} ~ .
\eea
$K_2$ is the squeezing generator \cite{Yuen:1976vy}.
Note that the ${\tilde K}$'s and $K$'s can be constructed by using convenient combinations of the  generators $K_{\si,i}~$, $\si = a,b~$; $~i=1,2,3$:
$K_{\si,1} = \frac{1}{4}(\si^2 + {\si^\dag}^2)$, $K_{\si,2} =
\frac{i}{4}(\si^2 - {\si^\dag}^2)$, $K_{\si,3}= \frac{1}{2}({\si^\dag} \si +
\frac{1}{2})$, which close the su(1,1) algebra for each $\si$. Other formulas used in the text are:
\bea \lab{costeta}
\sin \theta = \frac{2g}{A}~, ~\qquad \cos \theta = \frac{\om_{-}}{A}~, \hspace{0.3cm}&&\\
\sinh \al = - \frac{4g^2}{B}~, \quad \cosh \al = \frac{\om_{+}A}{B}~, && \lab{coshal}
\eea
which have been obtained from Eq.~(\ref{teta}) and where
\bea \lab{Ad} A \equiv {\sqrt{\om_{-}^{2} + 4g^2}}~ , \quad
\lab{A} B \equiv \sqrt{\om_{+}^{2}A^{2} - 16g^{4}}~,
\eea
with $B$ assumed to be different from zero: $B \neq 0$. Moreover, we have also used
\be \lab{coms1}
[ J_2 , I_1  ] = i {\tilde K}_{1} ~ , \quad  [J_2 ,K_2 ] = 0 ~ ,
\quad [K_2 ,J_{3}] = i {\tilde K}_{1} ~ ,
\ee
\be
\lab{coms2}
[ K_{2} , I_{3} ] = i K_1 ~ , \quad [ K_2 ,{\tilde K}_{1}] = i J_3 ~ ,
\ee
and obtained
\bea   \lab{IHteta}
H (\theta) = e^{i \theta I_{1}} H~ e^{-i \theta I_{1}} \hspace{3.7cm}&& \\
= \om_+ (J_3 - \frac{1}{2}) + A~  I_3  - \frac{2g \om_{-}}{A} ~J_2 \hspace{1.15cm}&& \non \\
- \frac{4g^{2}}{A} ~{\tilde K}_1~, \hspace{3.9cm}&&  \non
\\ \non \\
\lab{KH}
H(\theta,\al) = e^{i \al K_{2}} H(\theta)~e^{-i \al K_{2}} \hspace{2.66cm} && \\ \non
= \frac{B}{A}  J_3 - \frac{1}{2}\om_{+} + \frac{\om_{+}{A^{2}}}{B} ~I_3 -  \frac{2g\om_{-}}{A} ~J_2 && \\
  + 4g^2\frac{A}{B} ~K_1~ . \hspace{3.3cm}&& \non
\eea

\end{document}